\documentclass[showpacs,preprintnumbers,amsmath,amssymb]{revtex4}
\usepackage{amsmath}
\usepackage{latexsym}
\usepackage{bbm}
\usepackage{graphicx}
\usepackage{psfrag}

\begin{document}

\title{Tests of the lattice index theorem}

\author{Gerald Jordan}
 \altaffiliation[Also at ]{Photonics, Vienna University of Technology, Gusshausstrasse 25-29, 1040 Vienna, Austria.}
\author{Roman H\"ollwieser}
\author{Manfried Faber}
\affiliation{Atomic Institute, Vienna University of Technology, Wiedner Hauptstr. 8-10, 1040 Vienna, Austria}
\email{faber@kph.tuwien.ac.at}
\author{Urs M. Heller}
\affiliation{American Physical Society, One Research Road, Box 9000, Ridge, NY 11961-9000, USA}

\date{\today}

\begin{abstract}
We investigate the lattice index theorem and the localization of the zero-modes for thick classical center vortices. For non-orientable spherical vortices, the index of the overlap Dirac operator differs from the topological charge although the traces of the plaquettes deviate only by a maximum of 1.5\% from trivial plaquettes. This may be related to the fact that even in Landau gauge some links of these configuration are close to the non-trivial center elements.
\end{abstract}

%{\bf Keywords:}center vortices, Atiyah-Singer index theorem, overlap operator, lattice gauge theory

\maketitle

\section{Introduction}
%----------------------------------------------------------------------
% INTRODUCTION
%----------------------------------------------------------------------
At large separations static color sources of non-vanishing N-ality 
give rise to gluonic flux tubes and a linearly rising potential. This 
behavior is caused by the non-perturbative QCD-vacuum which compresses 
the gluonic flux. Center vortices 
\cite{tHooft:1977hy,Vinciarelli:1978kp,Yoneya:1978dt,Cornwall:1979hz,Mack:1978rq,Nielsen:1979xu},
quantized magnetic flux lines, play an important role for the confinement of color charges,
 as has been shown by numerical calculations\cite{DelDebbio:1996mh,Kovacs:1998xm}.
 In addition, simulations have indicated that vortices could also account 
for phenomena related to chiral symmetry, such as topological charge and 
spontaneous chiral symmetry breaking (SCSB)
\cite{Reinhardt:2000ck,deForcrand:1999ms,Alexandrou:1999vx,Engelhardt:2002qs}.

It is well-known that these non-perturbative features of QCD are intimately 
linked to the properties of the low-lying spectrum of the Dirac operator. 
The topological charge of a gauge field equals the index of the Dirac operator, 
while the chiral condensate, the order parameter for SCSB, is proportional 
to the spectral density of the near-zero modes.

With the advent of the overlap operator \cite{Narayanan:1994gw,Neuberger:1997fp},
the fundamental problems of investigating chiral features on the
lattice have been overcome. The overlap operator has
an exact chiral symmetry \cite{Luscher:1998pqa}, implements a lattice version
of the index theorem \cite{Narayanan:1994gw}, and may even be used for the 
definition of a local topological charge density \cite{Reinhardt:2000ck}.

In this paper, we report on our calculations with the overlap operator
applied to thick classical center vortices. We investigate the localization 
of zero-modes with respect to the position of the thick vortices and find
an interesting discrepancy in the topological charge determined by
different methods.

\section{Strategy}
%----------------------------------------------------------------------
% TECHNICAL DETAILS
%----------------------------------------------------------------------

% VORTICES
We work with SU(2) gauge fields on lattices with $N_s$ sites in the x-, 
y- and z-direction, and $N_t$ sites in the t-direction.  In four 
dimensions vortices form closed two-dimensional surfaces. We investigate 
thick classical center vortices in the shapes of planes (closed by 
lattice periodicity) and spheres. The details of the individual vortex 
types will be discussed along with our results.

% TOPOLOGICAL CHARGE
We compare different definitions of the lattice topological charge: 
\begin{itemize}
\item[(1)] The topological charge determined from the shape of P-vortices, 
that means from vortex intersections and writhing points according to 
ref.~\cite{Reinhardt:2000ck}.

\item[(2)] The index of the overlap Dirac operator \cite{Narayanan:1994gw,Neuberger:1997fp}.

According to the Atiyah-Singer index theorem, which holds for continuous
fields, the topological charge is given by the index
\begin{equation}
    \mathrm{ind} D[A] = n_- - n_+ = Q
\label{eq:index}
\end{equation}
where  $n_-$ and $n_+$ are the number of left- and right-handed zero modes 
\cite{Atiyah:1971rm,Schwarz:1977az,Brown:1977bj}.
It has been shown that this relation is also valid on the lattice 
for the overlap Dirac operator: its index coincides with
the discretized gluonic definition in the continuum limit \cite{Adams:2000rn}.

The overlap Dirac operator is defined by \cite{Edwards:1998yw}
\begin{gather}
  D = \frac{1}{2} [ 1 + \gamma_5 \epsilon(H^+_L) ]
\label{eq:D_ov}
\end{gather}
Here, $\epsilon$ is the sign function,
\begin{gather}
  H_L^+ = \gamma_5 D_w(-m_0),
\end{gather}
and $D_w$ is the usual lattice Wilson Dirac operator with mass $-m_0$
(we use $m_0=1.0$).

The overlap operator can be used to define the topological charge only on
so-called ``admissible'' gauge fields. This restriction assures that
$H^+_L$ has no zero eigenvalues so that the sign-function is
well-defined. A sufficient, but not necessary condition for this is
\cite{Luscher:1998du,Neuberger:1999pz}
\begin{gather}
  \mathrm{tr}(\mathbbm 1 - U_{\mu\nu}) < 0.03
\label{eq:adm-cond}
\end{gather}

The diagonalization of the overlap operator yields the corresponding Dirac zero modes
with well-defined chirality properties. We further look at the distribution of the 
scalar fermionic density
\begin{equation}
\rho (x) = \sum_{c,d} |\vec v(x)_{cd}|^2,
\end{equation}
where the summation indices $c$ and $d$ refer to color and Dirac indices of 
the eigenvectors $\vec v$, respectively, and the left- or right-handed chiral densities
\begin{equation}
\rho_\pm (x) = \sum_{c,d,d^\prime} \vec v(x)_{cd}^*\frac{(1-\gamma_5)^{d,d^\prime}}{2}\vec v(x)_{cd^\prime}^2,
\end{equation}
where only zero-modes of the corresponding chirality contribute. When the zero modes of a specific chirality are degenerate, it is useful to investigate the square root of the sum of the scalar densities of these modes. This average density respects eventual symmetries of the field configuration. 

\item[(3)] The integral (sum, on the lattice) of the gluonic charge density
$q(x) = \frac{1}{16\pi^2} \, \mathrm{tr} (\mathcal F_{\mu\nu}
\tilde{\mathcal F}_{\mu\nu})$ 
in the ``plaquette'' and/or ``hypercube'' definitions on the lattice, 
see ref.~\cite{DiVecchia:1981qi,DiVecchia:1981hh}.
Since Monte-Carlo configurations are in general too coarse,
these definitions are usually only applied after cooling.

In the continuum, since the topological charge density $q(x)$ is the total 
derivative of the topological current $k_\mu$ \cite{vanBaal:1982ag},
\begin{eqnarray}
\begin{aligned}
q(x) &= \partial_\mu k_\mu,\\
k_\mu &= - \frac{1}{8\pi^2}\epsilon_{\mu\nu\rho\sigma}
 \mathrm{tr} \left[\mathcal A_\nu \partial_\rho\mathcal A_\sigma+
 \mathrm i\frac{2}{3}\mathcal A_\nu\mathcal A_\rho\mathcal A_\sigma\right],
  \label{eq:qk}
\end{aligned}
\end{eqnarray}
one can apply Gauss' theorem to transform the 
expression for $Q$ into a surface integral,
\begin{align}
  Q &=\int d^4x\;q(x)=\oint_{S^3}d\sigma_\mu k_\mu, \label{eq:Qsfc}
\end{align}
provided the gauge field is smooth.
Assuming a sufficiently fast decaying field strength for $x^2\rightarrow\infty$,
the gauge field at infinity can be written as a pure gauge
\begin{gather}
  \mathcal A_\mu(x^2\rightarrow\infty)=\mathrm i\partial_\mu\Omega\Omega^\dagger.
\end{gather}
It can be shown that the surface integral Eq.~(\ref{eq:Qsfc}), and hence $Q$ for smooth fields,
 measures the winding number of $\Omega$, which maps the group 
space SU(2) onto the space-time surface $S^3$ at infinity.

\end{itemize}

\section{Description of configurations and results}
%----------------------------------------------------------------------
% RESULTS + DISCUSSION
%----------------------------------------------------------------------

\subsection{Plane vortices}
% PLANE VORTICES
The construction of planar vortices is explained in \cite{Greensite:2003bk}. 
We use periodic boundary conditions where plane center vortices always 
come in pairs. All three definitions of $Q$ yield identical results 
for all configurations containing only plane vortices.

\begin{figure}
\psfrag{x}{$x$}\psfrag{y}{$y$}\psfrag{z}{$z$}
\psfrag{1}{$1$}\psfrag{6}{$6$}\psfrag{12}{$12$}
\begin{tabular}{ccc}
Geometry & Topological charge density & Fermionic density \\
\includegraphics[keepaspectratio,width=0.3\textwidth]{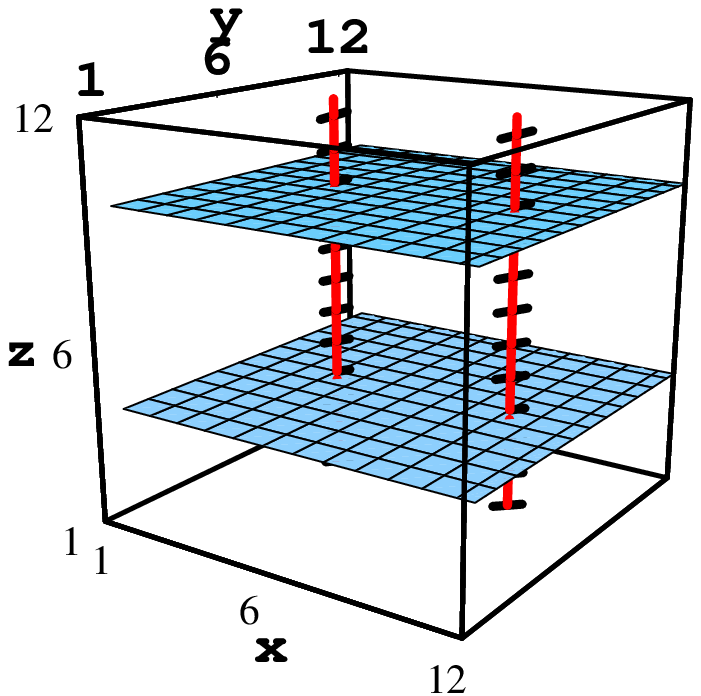} &
\includegraphics[keepaspectratio,width=0.3\textwidth]{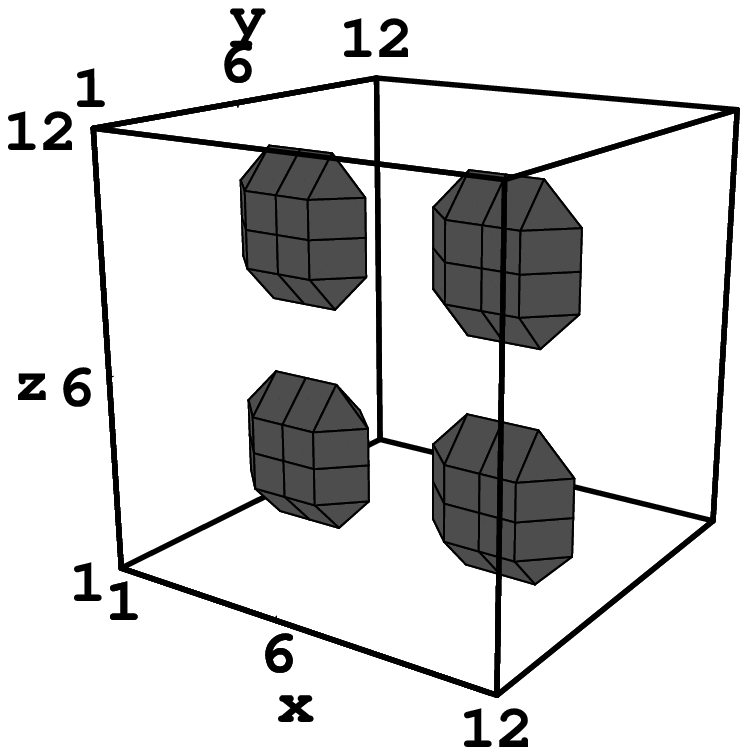} & 
\includegraphics[keepaspectratio,width=0.3\textwidth]{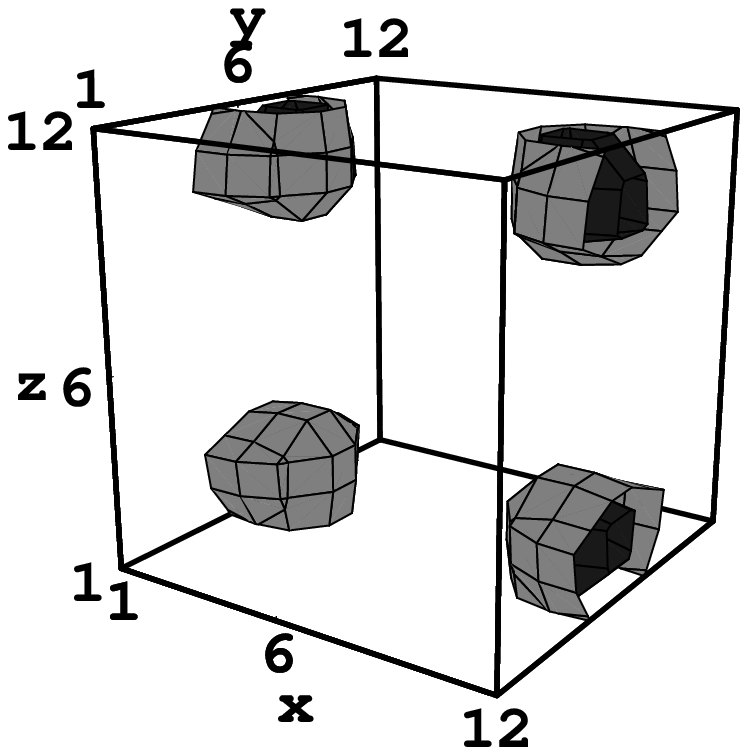}
\end{tabular}
\caption{Plane vortices on a $12^4$-lattice in xy- and zt-plane intersect 
in four points giving topological charge and fermionic density.}
\label{fig_plq2}
\end{figure}

As an example, we show in Fig.~\ref{fig_plq2} some diagrams for two 
orthogonal pairs of plane vortices on a $12^4$ lattice, which intersect 
in 4 places. The left picture shows the position of the P-plaquettes 
after going to maximal center gauge and center projection. One vortex pair extends in x-y-planes, the other pair in z-t-planes. The short pieces of lines 
attached to the vertical line symbolize the extent of the $z$-$t$-vortices in the time-direction. The diagram in the middle of Fig.~\ref{fig_plq2} shows an equi-density surface of the topological charge density determined in the plaquette definition and the right diagram shows an equi-density 
surface of the scalar fermionic density.

Each of the four intersection points gives rise to a lump of topological 
charge $Q=\pm\frac{1}{2}$\cite{Reinhardt:2000ck}. The sign of the contribution at 
a given intersection point can be changed by a flip of the orientation of 
the vortex surface, this means by a transition from links $U_\mu(x)$ creating 
one of the vortices to $U_\mu^\dagger(x)$ in the region of the intersection. 
After abelian projection a smooth transition of the vortex orientation in 
color space produces a monopole line. In this way one can produce 
configurations with two vortex pairs of topological charge $Q=0, \pm 1$ or $\pm 2$.
In all these cases the lattice index theorem is fulfilled and 
our results agree with the analytical solution for the zero modes presented 
in \cite{Reinhardt:2002cm}.

\subsection{A non-orientable spherical vortex}
\begin{figure}
\psfrag{linkphase}{link phase}
\psfrag{pi}{$\pi$}
\psfrag{pi/2}{$\pi/2$}
\psfrag{x}{$x$}
\psfrag{0}{$0$}
\psfrag{10}{$10$}
\psfrag{20}{$20$}
\psfrag{30}{$30$}
\psfrag{40}{$40$}
\centering
\begin{tabular}{cc}
  \begin{minipage}{0.4\linewidth}
    \centering
    \includegraphics[width=0.66\textwidth]{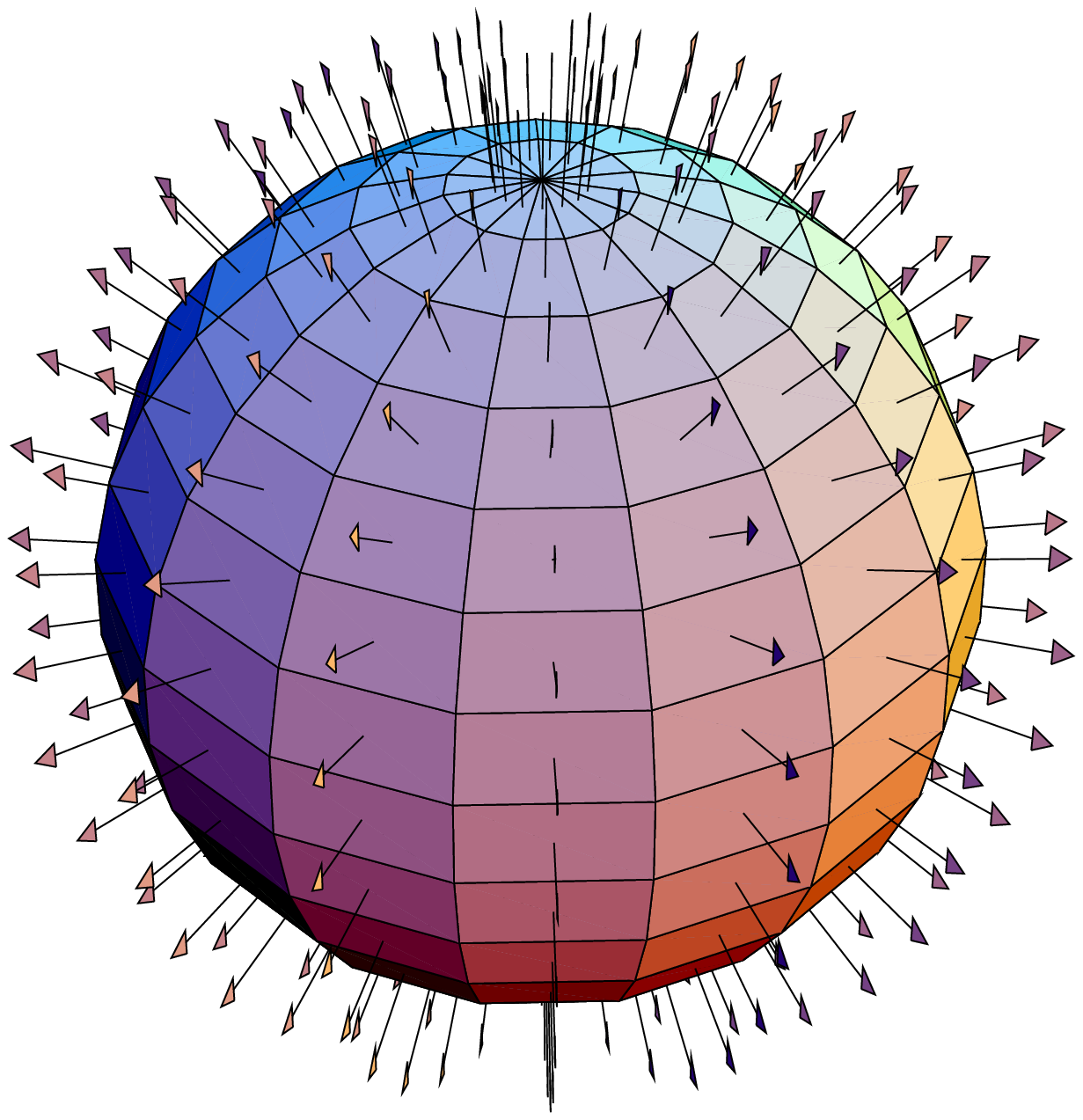}\\ \hspace*{0.8mm}\includegraphics[width=1.02\textwidth]{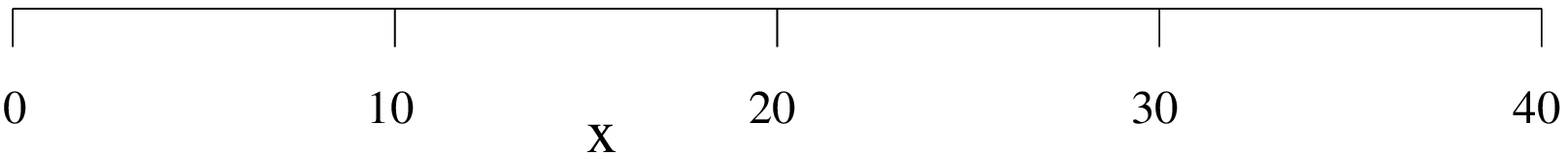} 
  \end{minipage}
& \includegraphics[width=0.5\textwidth]{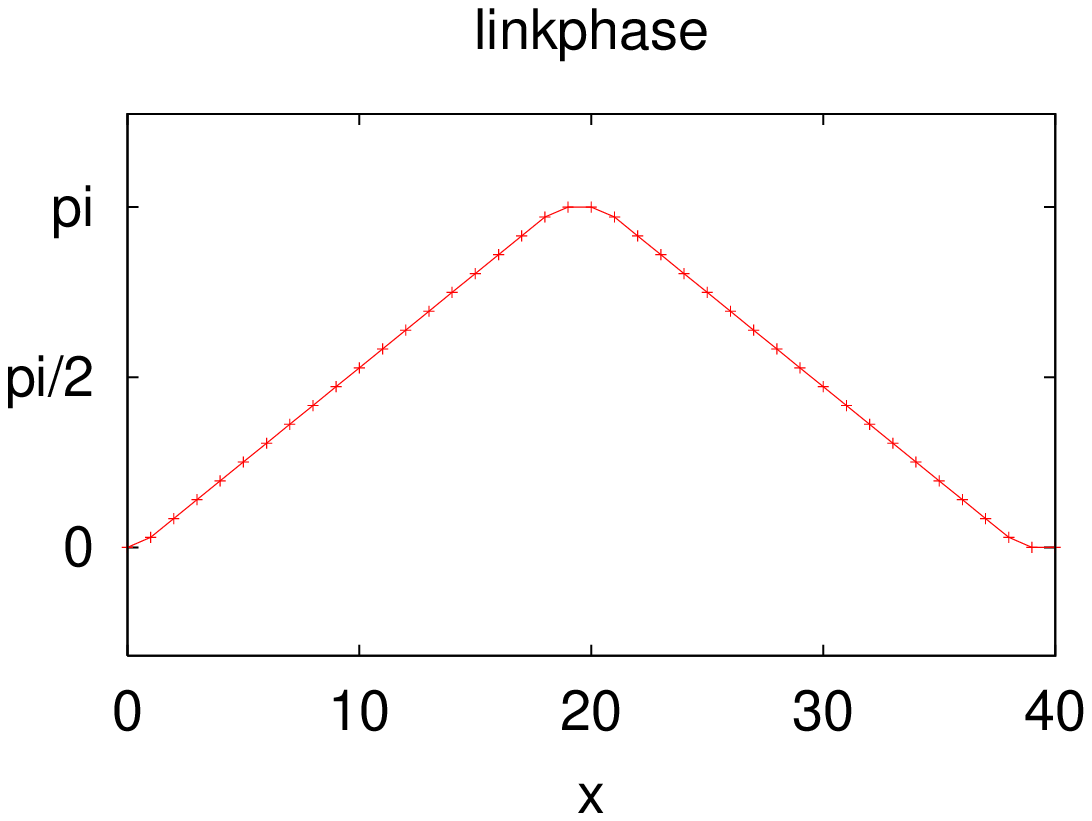} 
\end{tabular}
\caption{Thick Spherical SU(2)-vortex (hedgehog, non-orientable) and change of its link phase $\alpha(r)=\alpha_-(r)$ according to Eq.~(\ref{hedgehog}).}
\label{fig_sph-n}
\end{figure}

% SPHERICAL VORTICES
A non-orientable spherical vortex can be constructed with an appropriate 
``profile''-function $\alpha(|\vec r-\vec r_0|)$ by time-like links which form
a hedgehog in one time-slice of the lattice, where $\vec r_0$ is the 
center of a spatial sphere
\begin{gather}\label{hedgehog}
U_\mu(x) = \begin{cases}
 \exp\left\{\mathrm i\,\alpha(|\vec r-\vec r_0|)\,\vec n(x)\cdot\vec\sigma\right\}
 &t=1,\,\mu=4,\\
 \mathbbm{1}&\mathrm{elsewhere}.\end{cases}
\end{gather}
A characteristic property of a hedgehog configuration is the agreement 
between the color direction $\vec n$ and the spatial direction of $\vec r-\vec r_0$
\begin{gather}\label{ColorVectors}
\vec n(\vec r,t)=\frac{\vec r-\vec r_0}{|\vec r-\vec r_0|},
\end{gather}
see left diagram in Fig.~\ref{fig_sph-n}. For the choice (\ref{hedgehog}) of the 
links the time-like Wilson lines (holonomies) agree with the time-like links 
in the $t=1$ time-slice.

At $\vec r=\vec r_0$ and at the largest possible distances from $\vec r_0$  
on a periodic lattice the direction of the color-vectors (\ref{ColorVectors}) 
are undefined. This does not lead to a singularity of the gauge field if the 
profile-function $\alpha(r)$ is appropriately chosen such that the Wilson 
lines at these special points become center elements of the gauge group. 
A good choice is $\alpha(r)=\alpha_-(r)$ with
\begin{gather}
\alpha_-(r) = \begin{cases} \pi & r < R-\frac{\Delta}{2} \\
     \frac{\pi}{2}\left( 1-\frac{r-R}{\frac{\Delta}{2}} \right) & R-\frac{\Delta}{2} < r < R+\frac{\Delta}{2} \\
                        0 & R+\frac{\Delta}{2} < r.
          \end{cases}\label{minus}
\end{gather}
As shown in the right diagram of Fig.~\ref{fig_sph-n} for a $40^3\times 2$ 
lattice
the phase $\alpha$ of the links in the first time slice varies from $\pi$ at 
the center of the sphere to $0$ outside of the vortex core of thickness $\Delta$, chosen as $\Delta=18$.

\begin{figure}[h]
\psfrag{x}{$x$}
\psfrag{y}{$y$}
\psfrag{z}{$z$}
\centering
\includegraphics[keepaspectratio,width=0.3\textwidth]{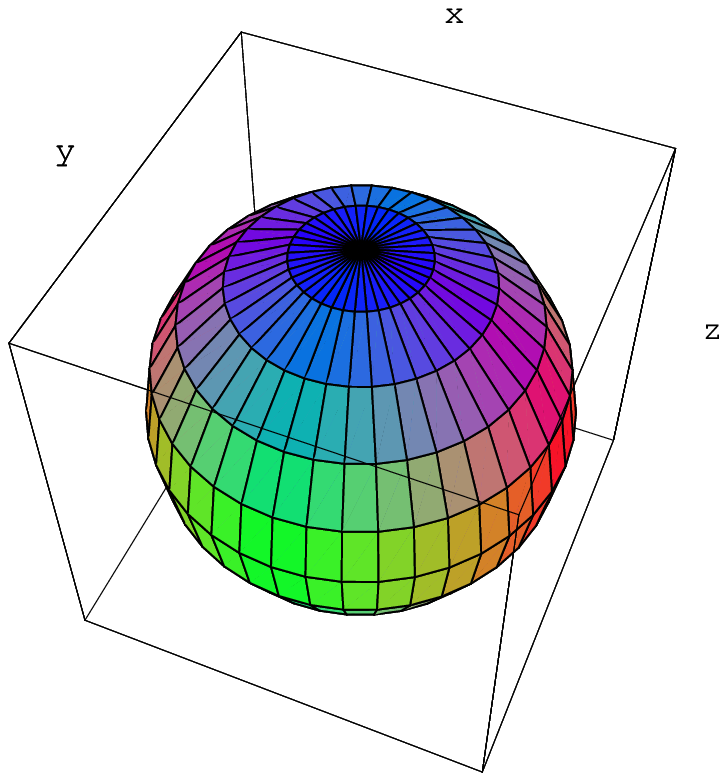} \hspace{10mm}
\includegraphics[keepaspectratio,width=0.3\textwidth]{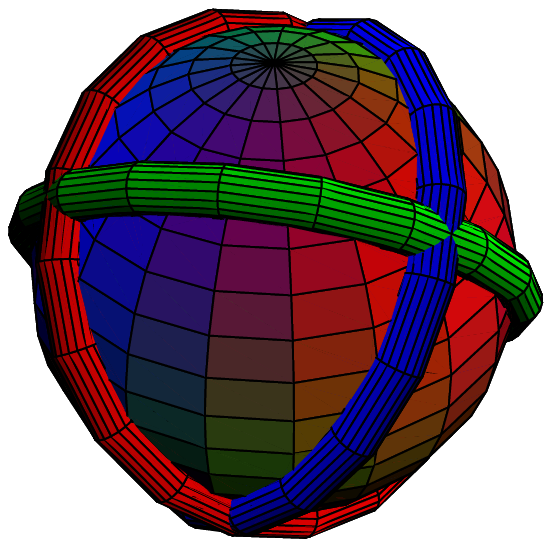}
\caption{The color structure of the vortex surface (l) leads to monopole lines 
after abelian projection (r).}
\label{fig_ab-proj}
\end{figure}

The check that this configuration is a vortex is done with maximal center gauge 
and center projection and results in a P-vortex forming a lattice representation 
of a sphere of radius $10$ in the first time slice. The color structure of the 
thick vortex which is symbolically indicated in the left diagram of 
Fig.~\ref{fig_ab-proj} leads to a monopole loop on a great circle of the P-vortex 
after maximal abelian gauge and abelian projection. The direction of the loop 
depends on the U(1) subgroup chosen as abelian degrees of freedom. For the subgroup 
defined by the Pauli matrices $\sigma_1, \sigma_2$ or $\sigma_3$ the monopole loops 
are in the y-z-, z-x- and x-y-plane, respectively. This is indicated schematically 
in Fig.~\ref{fig_ab-proj} with three colors.

The P-vortex surface which is a closed surface in dual space consists only of the 
duals of space-time plaquettes. There are no vortex intersections and no writhing 
points. Therefore, the topological charge $Q$ determined from the P-vortex surface 
\cite{Reinhardt:2000ck} vanishes, $Q=0$. The traces of all plaquettes are close to unity,
 $\mathrm{tr}(\mathbbm 1 - U_{\mu\nu}) \le 1-\cos\frac{\pi}{18}=0.015$, compare the 
link-phases in Fig.~\ref{fig_sph-n}. The topological charge in the plaquette 
definition or in the hypercube definition can be used safely without cooling. 
According to Eq.~(\ref{hedgehog}) the only non-trivial links are time-like, 
therefore all space-space plaquettes are trivial, which confirms the vanishing of the 
topological charge.

\begin{figure}
\centering
\psfrag{x}{$x$}
\psfrag{z}{$z$}
\psfrag{polyakovphase}{Polyakov phase}
\psfrag{pi}{$\pi$}
\psfrag{pi/2}{$\pi/2$}
\psfrag{0}{$0$}
\psfrag{10}{$10$}
\psfrag{20}{$20$}
\psfrag{30}{$30$}
\psfrag{40}{$40$}
\includegraphics[keepaspectratio,width=0.5\textwidth]{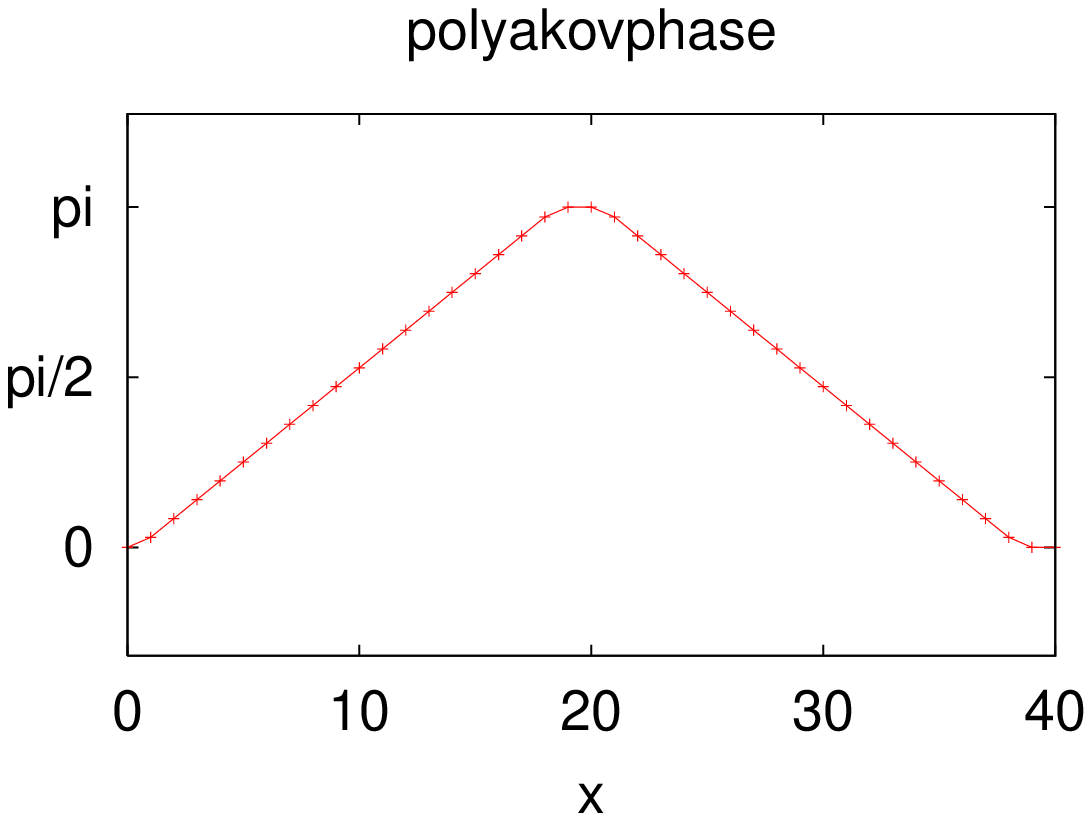}
\includegraphics[keepaspectratio,width=0.4\textwidth]{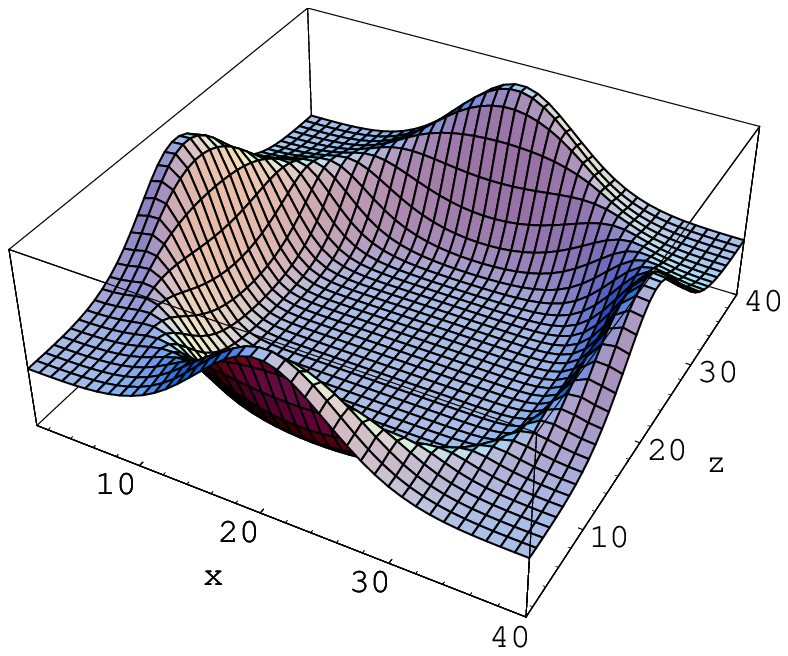}
\caption{Phases of Wilson lines along a line through the center of a spherical vortex (l), normalized scalar densities for all zero-modes of one chirality in the planes with $z=20$ close to the center of the sphere (r). The density vanishes in the region where the phase of the Polyakov loop is $\pi$. Both chiralities and both time-slices on the $40^3\times 2$-lattice give the same picture.}
\label{fig_sph-zerom}
\end{figure}

\begin{figure}
\psfrag{action  S}{action $S/S_\mathrm{inst}$}
\psfrag{2*40**3}{}
\psfrag{0}{$0$}
\psfrag{5}{$5$}
\psfrag{10}{$10$}
\psfrag{15}{$15$}
\psfrag{20}{$20$}
\psfrag{30}{$30$}
\psfrag{40}{$40$}
\psfrag{50}{$50$}
\psfrag{60}{$60$}
\psfrag{70}{$70$}
\psfrag{80}{$80$}
\psfrag{90}{$90$}
\psfrag{100}{$100$}
\psfrag{0.5}{$0.5$}
\psfrag{1}{$1$}
\psfrag{Q}{$Q$}
\psfrag{S}{$S$}
\psfrag{number of cooling steps}{number of cooling steps}
\psfrag{topological charge Q}{topological charge $Q$}
\centering
\includegraphics[keepaspectratio,width=0.5\textwidth]{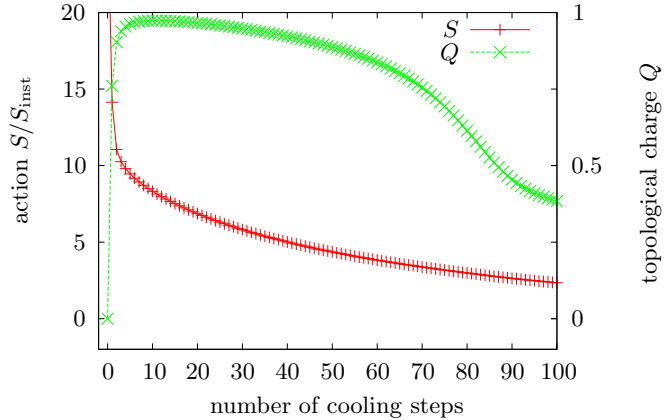}
\caption{By cooling the topological charge rises from zero close to one for $\alpha=\alpha_-$ (right scale) while the action $S$ (in units of the one-instanton action $S_{\rm inst}$) decreases slowly (left scale). The data is for a $40^3\times 2$-lattice.}
\label{fig_sph-cool}
\end{figure}

This configuration has three zero-modes \footnote{In this work, we denote as
``zero-mode'' any eigenmode of $D^\dagger D$ of the overlap Dirac operator
(\ref{eq:D_ov}) with eigenvalue smaller than $10^{-5}$. If such modes exist
of both positive and negative chirality they are most likely not exact
zero-modes, but lifted slightly above zero pairwise. This numerical
inaccuracy does not affect the value of the index (\ref{eq:index}).}
of positive chirality and four zero-modes 
of negative chirality. Normalized densities of positive and negative chirality modes 
have the same spatial distribution.\\  
For the spherical vortex with center at (19.5,19.5,19.5), we show the scalar 
density in the plane $z=20$ in Fig.~\ref{fig_sph-zerom}. The non-trivial links connect 
$z=20$ and $z=21$, and the densities in both time-slices are identical. 
The zero-modes are localized mainly in the region of trivial Wilson lines (for $|\vec r-\vec r_0|\sim 20$)
and seem compressed in the regions where the mirror pictures of the spherical vortex 
on the periodic lattice approach each other. In the region 
where the Wilson lines are non-trivial center elements (for $|\vec r-\vec r_0|$ around 0), the density vanishes.
According to the index theorem the topological charge is given by 
\begin{equation}
    Q = n_- - n_+ = 1.
\end{equation}
This value is in disagreement with the values determined above ($Q=0$). But it agrees with the gluonic result which one gets after cooling. The cooling history is shown in Fig.~\ref{fig_sph-cool}. The left scale shows the action $S=\sum_\Box(1-\frac{1}{2} \mathrm{Tr} U_\Box)$ in units of the one-instanton action $S_{\rm inst}$ and the right scale the value of the topological charge in the plaquette definition. The spherical vortex is not a minimum of the action, i.e. not an instanton. During cooling its action is first rapidly decreasing, whereas its topological charge starts with zero and reaches a value close to one after a few cooling steps. 
On the $40^3\times 2$-lattice the action does not really approach a plateau value. Since a plateau appears for a $12^4$-lattice this seems to be related to the strong asymmetry of the lattice. Hence one would expect to see a clear plateau on a symmetric, but finer $40^4$ lattice, as well. The number of zero-modes is the same for various lattice sizes we used: $12^4$, and $12^3\times 2$ up to $40^3\times 2$, increasing the spacial extent $L_s$ in steps of 4. We conclude that cooling and overlap fermions give a different topological charge than the discretized integral $\frac{1}{16\pi^2}\int d^4x\,\mathrm{tr} (\mathcal F_{\mu\nu}\tilde{\mathcal F}_{\mu\nu})$ even for very smooth field configurations.

The hedgehog configuration is characterized by links in the center of the spatial 
sphere which are non-trivial center elements and Wilson lines $L(\vec x)$, $\vec x=(x,y,z)$,
 with a topological winding number $\nu=1$,
\begin{equation}
\nu(\{L\})=\frac{1}{24\pi^2}\int\mathrm d^3r\epsilon_{ijk}\mathrm{Tr}(V_iV_jV_k),
\quad V_i=\partial_i LL^\dagger.
\label{PolWinding}
\end{equation}
The topological charge that we get in the cooling process and from the
index theorem agrees with this winding number. In the continuum limit,
the gauge field at the center of the hedgehog becomes singular. This
singularity is due to the topological obstruction and can't be removed
by a gauge transformation. With Landau-gauge, which we did for 100 gauge
copies on a $40^3\times 2$-lattice, it is not possible to get rid of links close 
to non-trivial center elements. For the lattice spacing $a$ approaching zero, 
the gauge field for these links, $\mathcal A_0(x)\approx \pi\sigma_3/(g a)$, 
diverges. In all gauge copies we found the same links with negative traces. 
The number of such negative links and the values of their traces are listed in Table~\ref{neglinksunorient}.

\begin{table}[h]
\begin{center}
\begin{tabular}{|c|ccccccc|}
\hline
number of links&8&24&24&32&48&24&48\\
\hline
$\tfrac{1}{2}\mathrm{tr}U_\mu(x)$&-0.93&-0.68&-0.46&-0.30&-0.18&-0.08&-0.01\\
\hline
\end{tabular}
\end{center}
\caption{Number of links with negative traces and trace values after Landau gauge for the non-orientable spherical vortex of Eq.~(\ref{hedgehog}) on a $40^3\times 2$-lattice, e.g. there are eight links with $\tfrac{1}{2}\mathrm{tr}U_\mu(x)=-0.93$.}
\label{neglinksunorient}
\end{table}

\subsection{Other configurations with spherical vortices}
% ORIENTABLE VORTEX
An orientable spherical vortex we construct with the prescription 
\begin{gather}\label{hedgehogorient}
U_\mu(x)=\begin{cases}
 \exp \left\{\mathrm i\,\alpha(|\vec r-\vec r_0|)\,|n_i(x)|\cdot\sigma_i\right\} &t=1,\mu=4,\\
 \mathbbm 1 & \mathrm{elsewhere}. \end{cases}
\end{gather}
Here, $|n_i(\vec r,t)|=\left|\frac{(\vec r-\vec r_0)_i}{|\vec r-\vec r_0|}\right|$ is 
the absolute value of the coordinates of the direction vector $\vec{n}$. 
These links cover only half of the SU(2) group space and this half they cover
twice. The time-like Wilson lines have topological winding number zero. This 
configuration has no zero-modes and all ways to determine the topological charge 
lead to the same result, $Q=0$. With maximal Abelian gauge no monopole lines appear. With Landau-gauge, which we did for 100 gauge copies on a $40^3\times 2$-lattice, one can get rid of links close to non-trivial center elements. The number of such negative links and the values of their traces are listed in Table~\ref{neglinksorient}.

\begin{table}[h]
\begin{center}
\begin{tabular}{|c|cccc|}
\hline
number of links&8&24&24&24\\
\hline
$\tfrac{1}{2}\mathrm{tr}U_\mu(x)$&-0.26&-0.15&-0.06&-0.008\\
\hline
\end{tabular}
\end{center}
\caption{Number of links with negative traces and trace values after Landau gauge for the orientable spherical vortex of Eq.~(\ref{hedgehogorient}) on a $40^3\times 2$-lattice.}
\label{neglinksorient}
\end{table}

\begin{figure}[h]
\centering
\psfrag{x}{$x$}
\psfrag{z}{$z$}
\psfrag{polyakovphase}{Polyakov phase}
\psfrag{pi}{$\pi$}
\psfrag{pi/2}{$\pi/2$}
\psfrag{0}{$0$}
\psfrag{10}{$10$}
\psfrag{20}{$20$}
\psfrag{30}{$30$}
\psfrag{40}{$40$}
\includegraphics[keepaspectratio,width=0.45\textwidth]{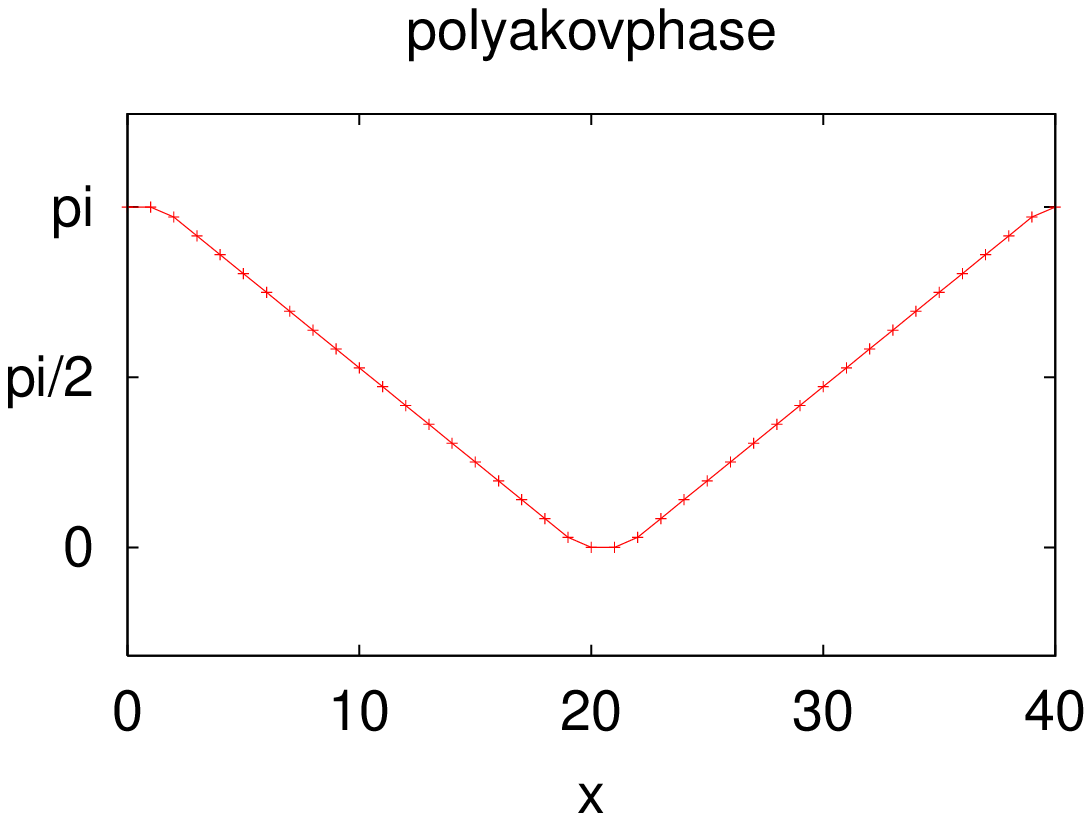}
\includegraphics[keepaspectratio,width=0.4\textwidth]{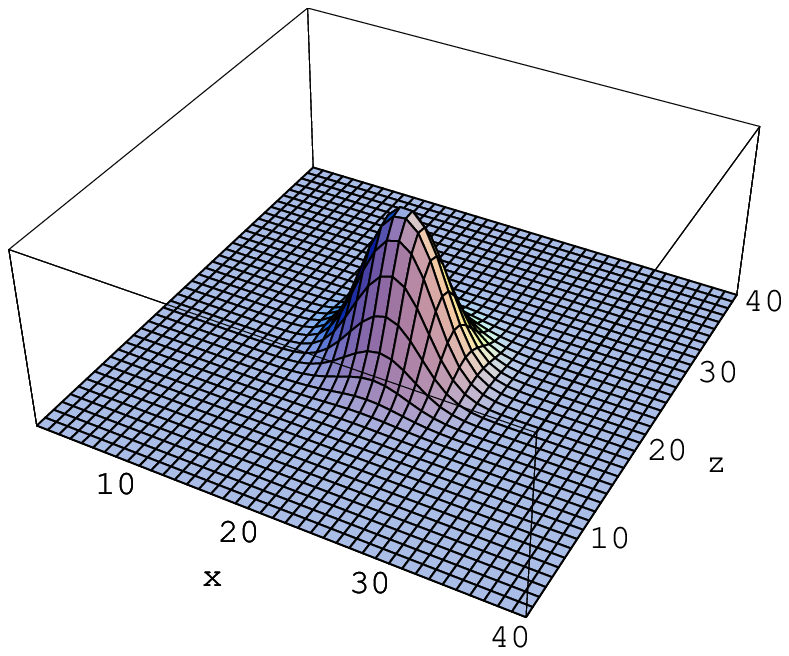}
\caption{Phase-diagram of the Polyakov loop (l) along a line through the center 
of a spherical vortex with the profile function $\alpha=\alpha_+$ (l) and scalar 
density of the only fermionic zero mode (r) which has positive chirality.}
\label{fig_sph-zerop}
\end{figure}

% OBSERVATIONS FOR SPHERES
A second type of a non-orientable spherical vortex we construct with the prescription 
(\ref{hedgehog}) and the profile function $\alpha(r)=\alpha_+(r)$,
\begin{gather}
\alpha_+(r) = \begin{cases} 0 & r < R-\frac{\Delta}{2} \\
                          \frac{\pi}{2}\left( 1+\frac{r-R}{\frac{\Delta}{2}} \right) & R-\frac{\Delta}{2} < r < R+\frac{\Delta}{2} \\
                          \pi & R+\frac{\Delta}{2} < r
          \end{cases}\label{plus}
\end{gather}
The phases of the links in the time slice agree again with the
phases of the Wilson lines in time direction and are shown in
Fig.~\ref{fig_sph-zerop}. In the center of the spherical vortex the
temporal Wilson lines are trivial and at ``infinity'' they are non-trivial
center elements. This configuration gives only one zero-mode of negative
chirality and none of positive chirality. The zero-mode is localized
in the center of the spatial sphere and in both time-slices occupied
by the vortex.  Again, the scalar density prefers regions of trivial
Wilson lines over those of non-trivial values. According to the index
theorem the topological charge is -1 and agrees with the winding number
(\ref{PolWinding}) of the hedgehog of Wilson lines. The topological
charge computed from plaquettes is $Q=0$ while the plaquette traces
differ by less than 1.5 \% from
trivial ones. Since this configuration is related to the configuration
in Fig.~\ref{fig_sph-zerom} by a non-periodic gauge transformation,
it is not surprising that again the different determinations of the
topological charge do not agree.

\begin{figure}[h]
\centering
\psfrag{x}{$x$}
\psfrag{x}{$x$}
\psfrag{z}{$z$}
\psfrag{10}{$10$}
\psfrag{20}{$20$}
\psfrag{30}{$30$}
\psfrag{40}{$40$}
\begin{tabular}{cc}
\includegraphics[width=0.4\columnwidth,keepaspectratio]{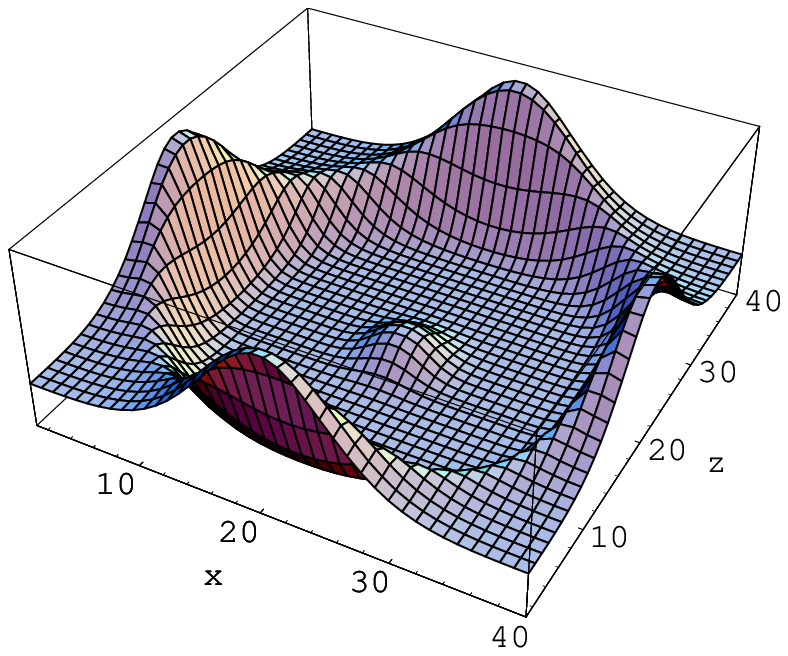} &
\includegraphics[width=0.4\columnwidth,keepaspectratio]{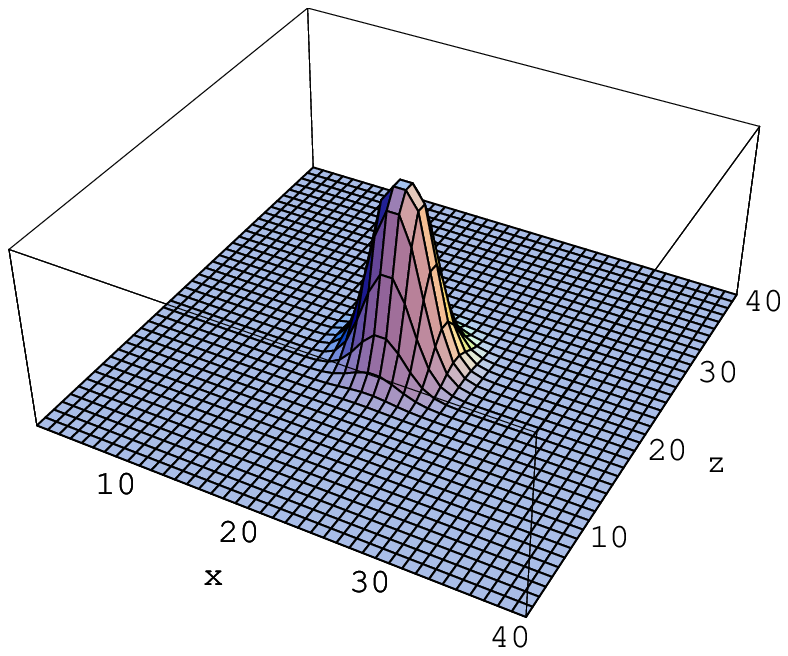} \\
$t=1$ & $t=2$
\end{tabular}
\caption{Scalar densities of fermionic zero modes of negative chirality for two equal vortices with the 
profile function $\alpha=\alpha_+$, in the time slices $t=1$ and $t=2$ of the 
$40^3\times 2$-lattice.}
\label{fig_twovort}
\end{figure}

For two equal vortices of this type, $\alpha=\alpha_+$, one in time slice $t=1$ 
and another at $t=2$ of the $40^3\times 2$-lattice we get three zero-modes of 
positive chirality with a scalar density like the diagram in Fig.~\ref{fig_sph-zerom} 
and five of negative chirality with the densities shown in Fig.~\ref{fig_twovort}. 

Since the modes are degenerate, they can mix and the plots display the sum of 
the densities of all zero-modes.
The left diagram displays the $t=1$ slice, where only four of these modes contribute significantly,
while the fifth mode leads only to the small peak in the middle.
Conversely, the density in the $t=2$ slice (shown in the right picture) is dominated by the fifth
mode in the center, the other modes being of a magnitude invisible on this scale.

The phase of the Polyakov line for this configuration is zero in the center of the 
spherical vortices and due to the $2\pi$-periodicity in the region between the 
mirror pictures on the periodic lattice. Again the zero-modes are localized in the 
region with trivial time-like Wilson lines.

A configuration with two vortices of different type in the two time-slices, one with 
$\alpha=\alpha_-$ at $t=1$ and another with  $\alpha=\alpha_+$ at $t=2$ does not 
show any zero-mode. In this case there is no discrepancy in the determinations of 
the topological charge. The phase of the Wilson line, which is the sum of the phases 
from the two vortices, is constant and equal to $\pi$.

A summary of the the number of zero-modes for the configurations described above
with spherical vortices is given in Table~\ref{sumspher}.
\begin{table}[h]
\begin{center}
\begin{tabular}{|c|cc|}
\hline
profile $\alpha$&$n_+$&$n_-$\\
\hline
$\alpha_-$ at $t=1$&3&4\\
$\alpha_+$ at $t=1$&1&0\\
$\alpha_-$ at $t=1$ and $\alpha_+$ at $t=2$ &3&5\\
$\alpha_+$ at $t=1$ and $\alpha_-$ at $t=2$ &5&3\\
$\alpha_+$ at $t=1$ and $\alpha_+$ at $t=2$ &0&0\\
\hline
\end{tabular}
\end{center}
\caption{Number of zero-modes for various smooth field configurations with 
spherical vortices. The profile functions $\alpha_\pm$ are defined in 
Eq.~(\ref{minus}) and (\ref{plus}).}
\label{sumspher}
\end{table}

The above table can be summarized in the following empirical rule:
Spherical vortices in slices (3D volumes) of the lattice contribute to the 
Dirac operator index with an integer given by the winding number (\ref{PolWinding}) 
of the corresponding Wilson lines, mapping the 3D volume of the slice to the SU(2) 
manifold of the Wilson lines.

%----------------------------------------------------------------------
% CONCLUSION
%----------------------------------------------------------------------
\section{Conclusion}
We have investigated various configurations with plaquettes close to unity. 
For configurations with a non-zero winding number of the Wilson lines we get disagreement 
between the topological charge determined by the plaquette and hypercube definition, and 
from the vortex intersections and writhing points on one side, and 
the topological charge from the index theorem and cooling on the other side. The 
transition in Eq.~(\ref{eq:Qsfc}), from the 4D integral over the topological charge density, 
to the 3D integral of the SU(2)-valued map $\Omega$, is only possible 
for smooth gauge fields. Vortex configurations with non-trivial winding numbers have 
singularities in the gauge fields which can't be removed by gauge transformations or 
smooth deformations of the field configuration. According to the vortex model of 
confinement vortices form closed two-dimensional surfaces which in the confined phase 
penetrate the whole space-time. In ref.~\cite{Bertle:1999tw} it was shown that P-vortex 
surfaces are mostly non-orientable. It is very likely that this non-orientability is 
related to a non-zero winding number of the gauge field and the artificial configurations 
which we discussed in this article reflect properties of those field configuration which 
saturate the path integral. The discrepancies which we have indicated are not due to 
the coarseness of the lattice, the traces of all plaquettes are close to unity, 
$\mathrm{tr}(\mathbbm 1 - U_{\mu\nu}) \le 1-\cos\frac{\pi}{18}=0.015$ but probably 
due to singularities in the gauge field related to non-trivial mappings of Wilson lines.
 
\section*{Acknowledgment}
This study was partially supported by the Austria Science Fund (FWF) under grant P20016-N16.
\bibliographystyle{h-physrev}
\bibliography{literatur}

\begin{thebibliography}{10}

\bibitem{tHooft:1977hy}
G.~'t~Hooft,
\newblock Nucl. Phys. {\bf B138}, 1 (1978).

\bibitem{Vinciarelli:1978kp}
P.~Vinciarelli,
\newblock Phys. Lett. {\bf B78}, 485 (1978).

\bibitem{Yoneya:1978dt}
T.~Yoneya,
\newblock Nucl. Phys. {\bf B144}, 195 (1978).

\bibitem{Cornwall:1979hz}
J.~M. Cornwall,
\newblock Nucl. Phys. {\bf B157}, 392 (1979).

\bibitem{Mack:1978rq}
G.~Mack and V.~B. Petkova,
\newblock Ann. Phys. {\bf 123}, 442 (1979).

\bibitem{Nielsen:1979xu}
H.~B. Nielsen and P.~Olesen,
\newblock Nucl. Phys. {\bf B160}, 380 (1979).

\bibitem{DelDebbio:1996mh}
L.~Del~Debbio, M.~Faber, J.~Greensite, and S.~Olejnik,
\newblock Phys. Rev. {\bf D55}, 2298 (1997), hep-lat/9610005.

\bibitem{Kovacs:1998xm}
T.~G. Kovacs and E.~T. Tomboulis,
\newblock Phys. Rev. {\bf D57}, 4054 (1998), hep-lat/9711009.

\bibitem{Reinhardt:2000ck}
H.~Reinhardt and M.~Engelhardt,
\newblock Center vortices in continuum yang-mills theory,
\newblock in {\em Quark Confinement and the Hadron Spectrum IV}, edited by
  W.~Lucha and K.~M. Maung, pp. 150--162, World Scientific, 2002,
  hep-th/0010031.

\bibitem{deForcrand:1999ms}
P.~de~Forcrand and M.~D'Elia,
\newblock Phys. Rev. Lett. {\bf 82}, 4582 (1999), hep-lat/9901020.

\bibitem{Alexandrou:1999vx}
C.~Alexandrou, P.~de~Forcrand, and M.~D'Elia,
\newblock Nucl. Phys. {\bf A663}, 1031 (2000), hep-lat/9909005.

\bibitem{Engelhardt:2002qs}
M.~Engelhardt,
\newblock Nucl. Phys. {\bf B638}, 81 (2002), hep-lat/0204002.

\bibitem{Narayanan:1994gw}
R.~Narayanan and H.~Neuberger,
\newblock Nucl. Phys. {\bf B443}, 305 (1995), hep-th/9411108.

\bibitem{Neuberger:1997fp}
H.~Neuberger,
\newblock Phys. Lett. {\bf B417}, 141 (1998), hep-lat/9707022.

\bibitem{Luscher:1998pqa}
M.~Luscher,
\newblock Phys. Lett. {\bf B428}, 342 (1998), hep-lat/9802011.

\bibitem{Atiyah:1971rm}
M.~F. Atiyah and I.~M. Singer,
\newblock Annals Math. {\bf 93}, 139 (1971).

\bibitem{Schwarz:1977az}
A.~S. Schwarz,
\newblock Phys. Lett. {\bf B67}, 172 (1977).

\bibitem{Brown:1977bj}
L.~S. Brown, R.~D. Carlitz, and C.-k. Lee,
\newblock Phys. Rev. {\bf D16}, 417 (1977).

\bibitem{Adams:2000rn}
D.~H. Adams,
\newblock J. Math. Phys. {\bf 42}, 5522 (2001), hep-lat/0009026.

\bibitem{Edwards:1998yw}
R.~G. Edwards, U.~M. Heller, and R.~Narayanan,
\newblock Nucl. Phys. {\bf B540}, 457 (1999), hep-lat/9807017.

\bibitem{Luscher:1998du}
M.~Luscher,
\newblock Nucl. Phys. {\bf B549}, 295 (1999), hep-lat/9811032.

\bibitem{Neuberger:1999pz}
H.~Neuberger,
\newblock Phys. Rev. {\bf D61}, 085015 (2000), hep-lat/9911004.

\bibitem{DiVecchia:1981qi}
P.~Di~Vecchia, K.~Fabricius, G.~C. Rossi, and G.~Veneziano,
\newblock Nucl. Phys. {\bf B192}, 392 (1981).

\bibitem{DiVecchia:1981hh}
P.~Di~Vecchia, K.~Fabricius, G.~C. Rossi, and G.~Veneziano,
\newblock Phys. Lett. {\bf B108}, 323 (1982).

\bibitem{vanBaal:1982ag}
P.~van Baal,
\newblock Commun. Math. Phys. {\bf 85}, 529 (1982).

\bibitem{Greensite:2003bk}
J.~Greensite,
\newblock Prog. Part. Nucl. Phys. {\bf 51}, 1 (2003), hep-lat/0301023.

\bibitem{Reinhardt:2002cm}
H.~Reinhardt, O.~Schroeder, T.~Tok, and V.~C. Zhukovsky,
\newblock Phys. Rev. {\bf D66}, 085004 (2002), hep-th/0203012.

\bibitem{Bertle:1999tw}
R.~Bertle, M.~Faber, J.~Greensite, and S.~Olejnik,
\newblock JHEP {\bf 03}, 019 (1999), hep-lat/9903023.

\end{thebibliography}

\end{document}